# Breakdown Current Density of Graphene Nano Ribbons

*Raghunath Murali\*, Yinxiao Yang, Kevin Brenner, Thomas Beck, and James D. Meindl*

Microelectronics Research Center, Georgia Institute of Technology, Atlanta GA

Graphene nanoribbons (GNRs) with widths down to 16 nm have been characterized for their current-carrying capacity. It is found that GNRs exhibit an impressive breakdown current density, on the order of $10^8$ A/cm$^2$. The breakdown current density is found to have a reciprocal relationship to GNR resistivity and the data fit points to Joule heating as the likely mechanism of breakdown. The superior current-carrying capacity of GNRs will be valuable for their application in on-chip electrical interconnects. The thermal conductivity of sub-20 nm graphene ribbons is found to be more than 1000 W/m-K.

\* raghu@gatech.edu, Ph: 404 385 6463

Graphene is a promising electronic material because of many interesting properties like ballistic transport[1], high intrinsic mobility[2], and width-dependent bandgap[3]. Graphene, in its 2D form, has been shown to have a high thermal conductivity[4] of around 5000 W/m-K pointing to its potential use as an on-chip heat spreader. Graphene nano ribbons (GNRs) have been predicted to be superior to Cu in terms of resistance per unit length[5] for use as on-chip interconnects. A high current-carrying capacity is critical for interconnect applications and reliability. There have been a number of studies on carbon nanotube (CNT) breakdown current density, and the current-carrying capacity of single-walled CNTs[6] is found to be on the order of $10^8$ A/cm$^2$; in carbon nanofibers, the breakdown current density ($J_{BR}$) has been measured[7] to be around $5 \times 10^6$ A/cm$^2$. Electrical breakdown has been used to burn away successive shells in a multi-wall CNT[8,9]. More recently, electrical breakdown has been used to obtain semiconducting CNTs from a mixture of CNTs since metallic ones burn away at a lower breakdown voltage[10]. Theoretical projections suggest that $J_{BR}$ of graphene should be on the same order as for CNTs. However, little experimental evidence exists on the electrical breakdown of either 2D graphene or 1D GNRs. In this work, it is experimentally shown that GNRs demonstrate an impressive $J_{BR}$. A simple relation between $J_{BR}$ and nanowire resistivity is seen to emerge from the experimental data.

Few-layer graphene (1-5 layers) is used as the starting material (see supporting material[11]). Each device consists of parallel ribbons fabricated between sets of electrodes, Fig. 1. The ribbon width between a pair of electrodes is designed to be the same for all the parallel ribbons. The range of widths studied in this work is 16nm<W<52nm, while the range of length is 0.2μm<L<1.0μm. Twenty one devices have been studied in this work, with each device yielding 5-10 GNRs (depending on the overlap of patterned channels to few-layer graphene) between the middle electrode pair. The outer electrode pair is used to test for contact resistance (in a four-point probe setup). A semiconductor analyzer is used to apply a voltage ramp (at the rate of 50 mV/s) between the middle electrodes. Due to increasing current density in the GNRs, there is a voltage at which a GNR breaks down, resulting in a visible drop in current. The device testing is stopped at this point, and low-bias measurements (for back-gated resistance and contact resistance) are made. The voltage ramp is then repeated from 0V. Successive GNR breakdowns occur at around the same voltage as for the first breakdown event. The breakdown current density of a GNR is extracted from the breakdown voltage and the resistance of the GNR; the resistance of a GNR for $J_{BR}$ calculation is extracted from the difference in conductance immediately before and immediately after a breakdown event.

Fig. 2a shows the I-V behavior of a device with 10 GNRs in parallel, and with W=22 nm, and L=0.75 μm. The I-V curves are for a set of parallel GNRs – the top I-V curve is for 10 GNRs in parallel, the second curve from top for 9 GNRs in parallel and so on. The I-V curve is initially linear, and becomes saturated at increasing bias. This saturation is repeatable as the sample is cycled from 0 V to 1.5 V; the contact resistance is found to be unchanged after bias-cycling. This indicates that the non-linearity at high-bias is due to self-heating effects and not due to contact annealing. Such I-V saturation has been observed at high-bias[12] in CNTs. Of the 21 devices tested in this work for breakdown current, 14 of the devices showed about a 2X increase in resistance (from low bias to the first breakdown event), 6 devices showed a 10-20% increase in resistance, whereas one device showed no increase in resistance. The reason for this varying behavior could be two fold: (i) varying impurity density between devices; the impurity density is estimated using the Dirac point shift[13] after contact metallization and is in the range $2-19\times10^{11}$ cm$^{-2}$ – a higher impurity density would cause more current saturation due to increased electron-phonon scattering, (ii) ballistic transport in short-length devices; it has been argued before[14] that ballistic transport (in CNTs) results in a linear I-V behavior with no current saturation at high bias. There are ten breakdown events for the device shown in Fig. 2a, corresponding to the ten GNRs in the device. It is found from repeated low-bias measurements immediately after a breakdown event that 2 to 3 min. is needed for a device to come back to its stable state from the self-heated state. Thus, low-bias measurements reported in this work are done 3 min. after any previous high-bias cycling. Fig. 2b shows the contact resistance after each breakdown event; the contact resistance is found to be almost constant after each event, and is usually in the range of 30-80 Ω for the devices reported in this work. With a contact area of 0.5-1 μm$^2$, this translates to a contact resistance of 15-80 Ω–μm$^2$. The breakdown voltage ($V_{BR}$) is seen to be around the same for all the ten GNRs in this device. Occasionally, it is seen that $V_{BR}$ of a later breakdown event is smaller than that of the previous event. This may occur if the device has not fully reached its stable state from the previous high-current cycle. Fig. 2c shows the breakdown current density of the ten GNRs – the range of current density is between 1.2 to 2.8 x$10^8$ A/cm$^2$; the variation in current density could be because of a variation in the number of layers or impurity density variation.

Fig. 3a shows breakdown current density of more than 100 GNRs versus their corresponding low-bias resistivity. A reciprocal relation is clearly seen between breakdown current density and nanowire resistivity ($\rho$). The best-fit for the data is obtained using the relation $J_{BR}=A\rho^{-B}$, where A=5.72x$10^8$, and B=0.71, with $\rho$ having the units of $\mu\Omega$-cm; the $R^2$ for this fit is found to be 74.4%. Note that $J_{BR}$ is extracted when the GNR is self-heated; the low-bias resistivity of a GNR, on the other hand, is extracted from the conductance difference between low-bias measurements done before and after a breakdown event. Fig. 3b shows $J_{BR}$ versus the high-bias resistivity (i.e. resistivity extracted from the conductance difference before and after a breakdown event). The best-fit for the data is again obtained using the relation $J_{BR}=A\rho^{-B}$, where A=9.57x$10^8$, and B=0.71; the $R^2$ for this fit is found to be 86.2%. Using the 1-D heat transport equation, a relation of the type $J_{BR} \propto 1/\sqrt{\rho}$ has been proposed[15]. The exponent of 0.71 extracted from the data suggests a faster breakdown with increasing resistivity; this indicates that the same factors that cause a higher resistivity also cause a degradation in breakdown current density – e.g. in-plane defects. For longer lengths, a relation of the type $J_{BR} \propto 1/\sqrt{(a\rho)}$ has been proposed[7], where $a$ is the cross-sectional area. Using a subset of data from Fig. 3b that have L>0.5 um, we get a fit (not shown) using the relation $J_{BR}=A(\rho a)^{-B}$ with B=-0.55; the fit has an $R^2$ of 92%.

It is possible to estimate the peak temperature in a GNR by solving the 1-D heat transport equation[16]:

$$V_{BR}I_{BR}\left(1-\frac{1}{\cosh(L/2L_H)}\right) = gL(T_{max}-T_0) \quad (1)$$

where $V_{BR}$ and $I_{BR}$ are the breakdown voltage and current respectively, $g$ is thermal conductance of the GNR (to the substrate and top-resist), $L$ is GNR length, $T_{max}$ is peak temperature in the GNR, and $T_0$ is the contact electrode temperature. Here $L_H$ is the characteristic thermal healing length along the GNR and is given by $(ka/g)^{1/2}$ where $k$ is thermal conductivity of the GNR. The relation can be rewritten as:

$$g(T_{max} - T_0) = J_{BR}^2 \rho A \left(1 - \frac{1}{\cosh(L/2L_H)}\right) \quad (2)$$

For an example GNR, $J_{BR}=7\times10^8$ A/cm$^2$ and $\rho_{3D}=100$ μΩ-cm (Fig. 3b). To evaluate $T_{max}$, it is necessary to assume values for g and k; from previously published results, g=0.20 W/m-k for bare CNTs on an oxide surface[16]. Previous measurements[17] on micron-wide, suspended graphene ribbons, at room temperature, yielded thermal conductivity values between 3080-5150 W/m-K. Since the GNR has a thin HSQ layer on the top, this contribution has to be included as well; thus the value of $g$ assumed above would need to be slightly higher than that found for bare CNTs on SiO$_2$. From (2), $T_{max}$ is found to be 180°C compared to 500-700 °C found for CNTs; if g is used as a fit parameter, even a low value of g=0.05 W/m-K results in a $T_{max}$ of only 195 °C. It is unlikely that GNRs would breakdown at such low temperatures – indeed, it has been recently reported[18] that the peak temperature in the middle of a micron-wide single-layer graphene on SiO$_2$ is more than 700 °C. Thus, k is used as a fit parameter to obtain realistic values of $T_{max}$. For k=1100 W/m-K, 0.15<g<0.30 W/m-K results in a $T_{max}$ between 700-800°C. The thermal conductivity thus extracted – 1100 W/m-K – is for an 18 nm wide ribbon. Similar calculations result in a thermal conductivity of 1000-1400 W/m-K for other GNRs. Edge roughness scattering of phonons in graphene ribbons has been argued to result in a size-dependent thermal conductivity[19]; it is found that k at room temperature reduces from 5500 W/m-K to 3000 W/m-K as the width of a single-layer graphene ribbon is scaled from 9 μm to 3 μm. In addition, umklapp scattering[19] too reduces k as the temperature of a graphene ribbon is increased beyond about 100 K. Since the GNRs under discussion are both narrow and self-heated to temperatures of 700-800 °C, it is expected that both edge roughness scattering and umklapp scattering would play an important role in determining thermal conductivity.

In conclusion, GNRs are found to display an impressive current-carrying capacity of more than $10^8$ A/cm$^2$, for widths down to 16 nm. The breakdown current density is found to have a reciprocal relationship to the nanowire resistivity, and points to Joule heating as the likely mechanism of breakdown. The extracted thermal conductivity of sub-20 nm GNRs is more than 1000 W/m-K.

Acknowledgements

The authors acknowledge funding support from Semiconductor Research Corporation/DARPA through the Interconnect Focus Center (IFC) and from Nanoelectronics Research Initiative (NRI) through the Institute for Nanoelectronics Discovery and Exploration (INDEX).

REFERENCES

1. C. Berger, Z. M. Song, X. B. Li, X. S. Wu, N. Brown, C. Naud, D. Mayou, T. B. Li, J. Hass, A. N. Marchenkov, E. H. Conrad, P. N. First, and W. A. de Heer, Science **312,** 1191-1196 (2006).
2. K. I. Bolotin, K. J. Sikes, Z. Jiang, M. Klima, G. Fudenberg, J. Hone, P. Kim, and H. L. Stormer, Solid State Communications **146,** 351-355 (2008).
3. M. Y. Han, B. Ozyilmaz, Y. B. Zhang, and P. Kim, Physical Review Letters **98,** 206805 (2007).
4. A. A. Balandin, S. Ghosh, W. Z. Bao, I. Calizo, D. Teweldebrhan, F. Miao, and C. N. Lau, Nano Letters **8,** 902-907 (2008).
5. A. Naeemi and J. D. Meindl, IEEE Electron Device Letters **28,** 428-431 (2007).
6. A. Javey, P. F. Qi, Q. Wang, and H. J. Dai, Proceedings of the National Academy of Sciences of the United States of America **101,** 13408-13410 (2004).
7. H. Kitsuki, T. Yamada, D. Fabris, J. R. Jameson, P. Wilhite, M. Suzuki, and C. Y. Yang, Applied Physics Letters **92,** 173110 (2008).
8. P. G. Collins, M. S. Arnold, and P. Avouris, Science **292,** 706-9 (2001).
9. M. Tsutsui, T. Yu-ki, S. Kurokawa, and A. Sakai, Journal of Applied Physics **100,** 94302 (2006).
10. B. Gyoung-Ho, H. Jea-Ho, J. Eun-Kyoung, S. Hye-Mi, O. L. Jeong, K. Ki-jeong, and C. Hyunju, IEEE Transactions on Nanotechnology **7,** 624-7 (2008).
11. See Supporting Material
12. P. G. Collins, M. Hersam, M. Arnold, R. Martel, and P. Avouris, Phys Rev Lett **86,** 3128-31 (2001).
13. Y. W. Tan, Y. Zhang, K. Bolotin, Y. Zhao, S. Adam, E. H. Hwang, S. Das Sarma, H. L. Stormer, and P. Kim, Physical Review Letters **99,** 246803 (2007).
14. P. Poncharal, C. Berger, Y. Yi, Z. L. Wang, and W. A. de Heer, Journal of Physical Chemistry B **106,** 12104-12118 (2002).
15. M. Suzuki, Y. Ominami, N. Quoc, C. Y. Yang, A. M. Cassell, and L. Jun, Journal of Applied Physics **101,** 114307 (2007).
16. E. Pop, D. A. Mann, K. E. Goodson, and H. J. Dai, Journal of Applied Physics **101,** - (2007).
17. S. Ghosh, I. Calizo, D. Teweldebrhan, E. P. Pokatilov, D. L. Nika, A. A. Balandin, W. Bao, F. Miao, and C. N. Lau, Applied Physics Letters **92,** 151911 (2008).
18. M. Freitag, M. Steiner, Y. Martin, V. Perebeinos, Z. Chen, J. C. Tsang, and P. Avouris, Nano Letters**,** Article ASAP.
19. D. L. Nika, E. P. Pokatilov, A. S. Askerov, and A. A. Balandin, Physical Review B (Condensed Matter and Materials Physics) **79,** 155413 (2009).

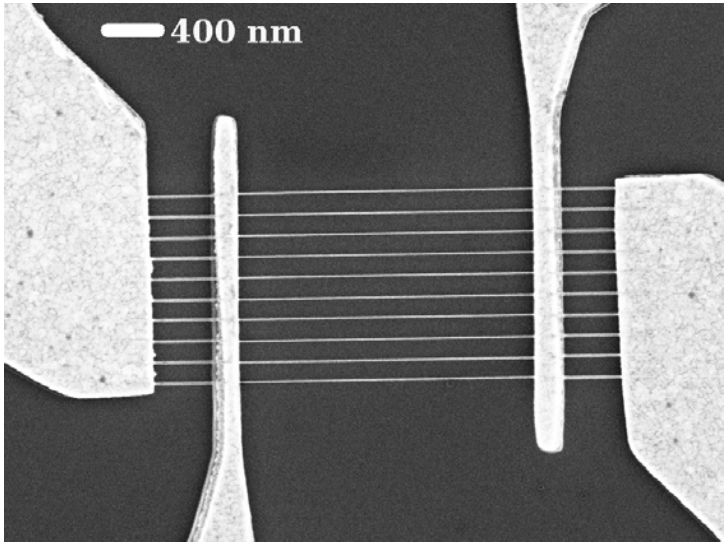

Figure 1: SEM image of a set of 10 GNRs between each electrode pair. The GNRs (below HSQ lines) are 21 nm wide between the middle electrode pair.

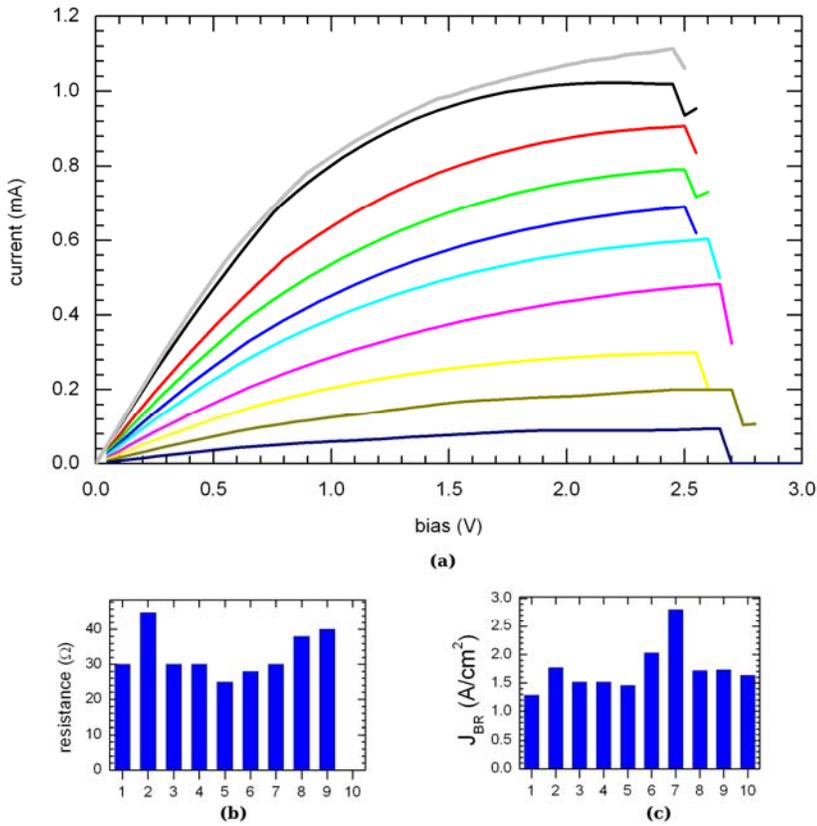

Figure 2: I-V curves of ten GNRs taken through electrical breakdown (a); each GNR has a width of 22 nm, and a length of 0.75 μm. The I-V curves are for a set of parallel GNRs – the top I-V curve is for 10 GNRs in parallel, the second curve from top for 9 GNRs in parallel and so on. The testing is stopped immediately after a breakdown event, followed by low-bias measurements of contact resistance (b). The breakdown current density of the ten GNRs is plotted in (c) with the units of $10^8$ A/cm$^2$.

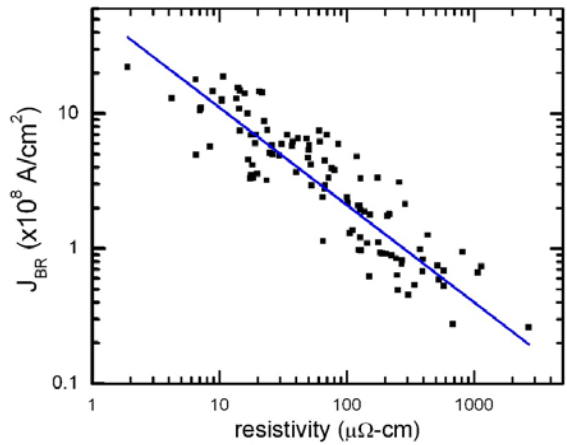

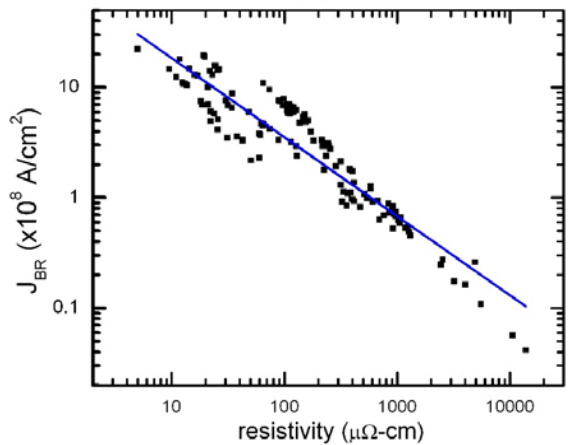

Figure 3: Breakdown current density versus resistivity; (a) shows a scatter plot with low-bias resistivity plotted on the x-axis – the $R^2$ for this fit is 74%; (b) shows a scatter plot for breakdown current density versus high-bias resistivity; the $R^2$ for this fit is 86%. The fit for both the plots is of the form $J_{BR}=A\rho^{-B}$ where B=0.71. If the breakdown mechanism was Joule heating, theory predicts that the exponent in the fit (B) should be 0.5; a steeper exponent in the fit indicates that the breakdown occurs faster for higher-resistivity GNRs, and might be indicative of higher defect densities contributing to faster electrical breakdown.

**Supporting Material**: Fabrication and characterization methods

Few layer graphene is used as the starting material. Graphene layers are flaked from large graphite pieces (Kish Graphite, Toshiba Ceramics) using a Scotch tape, and adhered onto an oxidized Si substrate with an oxide thickness of 300 nm. The Si substrate is degenerately doped for use as a back-gate. Using electron beam lithography, contact pads are fabricated by a metal lift-off process using Ti/Au (10nm/100nm) as the contact metal. This is followed by characterization of the 2D graphene for contact resistance, Dirac point, and carrier modulation. A second layer of lithography, using HSQ (XR-1541 resist, Dow Chemicals), defines nanometer-wide channels with dimensions in the range 16nm<W<52nm, and 0.2um<L<1.0um. A low-power oxygen plasma etch is used to transfer resist patterns into the channel. The XR-1541 resist stays on the GNR during the breakdown testing.

Electrical measurements were performed with standard lock-in techniques; excitation currents of 5nA-100nA were used in four-point probe measurements to extract contact resistance. A HP 4156 semiconductor parameter analyzer was used to perform low-bias measurements, along with a back-gate sweep. Tests for Ohmic contacts were performed at voltages down to a few micro-volts, and there was no indication of a Schottky barrier. High-resolution SEM imaging revealed the dimensions of the patterned GNRs; AFM scanning, confocal MicroRaman, and optical imaging were used to estimate the number of graphene layers and the quality of the layers.

All devices discussed thus far had HSQ covering the GNRs during the breakdown event. While HSQ provides an excellent pattern resolution, it is difficult to strip it so that the underlying GNR can be exposed; thus, when HSQ is used as the resist, it is not possible to image a GNR subjected to electrical breakdown. To enable removal of resist after GNR patterning, a different resist material (ZEP 520A, Zeon Chemicals) is used, resulting in a lower resolution (~100 nm) but successful removal of resist except for a 20 nm thick residue. Fig. S1-a shows a GNR patterned using this method; this GNR was subjected to electrical breakdown, and was imaged again, Fig. S1-b. The post-breakdown image clearly shows that breakdown occurs in the middle of the wire, pointing to Joule heating as the likely mechanism of electrical breakdown reported in this work. To compare against the data in Fig. 3b, the imaged GNR had a $J_{BR}$ of $1.2 \times 10^8$ A/cm$^2$ and a high-bias resistivity of 587 µΩ-cm.

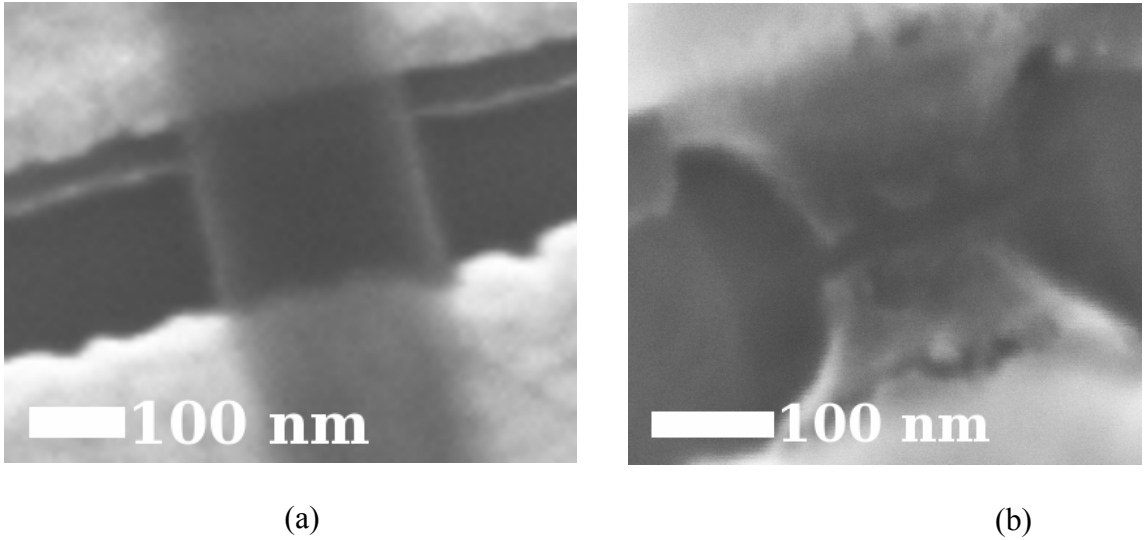

(a)          (b)

Figure S1: Before-and-after images of a 200 nm GNR, covered by 20 nm ZEP resist residue; (a) shows the intact GNR, and (b) shows the GNR after electrical breakdown. The breakdown is seen to occur in the middle, pointing to Joule heating as the likely mechanism of breakdown. This GNR had a breakdown current density of $1.2 \times 10^8$ A/cm$^2$ and a high-bias resistivity of 587 µΩ-cm.